\documentclass[journal=jacsat,manuscript=article]{achemso}

\usepackage{chemformula} 
\usepackage[T1]{fontenc} 
\usepackage[utf8]{inputenc}
\usepackage{upgreek}
\usepackage{amsmath}
\usepackage{lmodern}
\usepackage{amsfonts}
\usepackage{amsthm}

\usepackage{changes}
\usepackage{lipsum}

\author{Bidisha Bhatt}

\author{Shivam Gupta}

\author{Vasudevan Sumathi}

\author{Sivasurender Chandran}

\author{Krishnacharya Khare}

\email{kcharya@iitk.ac.in}

\affiliation{Department of Physics, Indian Institute of Technology Kanpur, Kanpur-208016, Uttar Pradesh, India}

\title{Reversible Dewetting of Thin Lubricating Films Underneath Aqueous Drops Using External Electric Field}

\begin{document}

\begin{abstract}
The stability of thin liquid films on a surface depends on the excess free energy of the system involving various short-range and long-range interactions. In an unstable condition, thin liquid films may dewet into multiple small-sized droplets via spinodal, homogeneous, or heterogeneous nucleation process. However, if the total excess free energy of the system can be manipulated using an external stimulus, one can control the stability of thin liquid films on demand. Here, we study the reversible dewetting process of thin lubricating films underneath aqueous drops on slippery surfaces using an external electric field. Upon applying voltage, stable thin lubricating films dewet in a manner identical to the spinodal dewetting of nanometer-thick liquid films. Upon removing the applied voltage, the dewetted droplets spread, coalesce with neighboring ones, and form a uniform film again, however the time taken during rewetting is very different compared to dewetting. The characteristic features of both the dewetting and rewetting processes are present over multiple cycles. Due to the random nature of the spinodal dewetting process, the final dewetting pattern does not show any correlation over multiple dewetting cycles.
\end{abstract}

\section{Introduction}
Solid surfaces coated with thin lubricating films are very common and desirable in many commercial and industrial applications, such as semiconductor device fabrication, friction reduction, heat transfer, protection, to name a few.\cite{erdemir2005review} In most of these applications, stability of the lubricating films determines the performance of the device. For thin lubricating films (thickness $\ll$ Capillary length), thin film stability can be defined in terms of the total free energy (per unit area) of the system due to the interfacial (short-range) and intermolecular (long-range) interactions. Under unstable condition, thin liquid films may rupture or dewet via different mechanism depending on the property of the system. Many research groups have investigated the deweting of thin liquid films to learn about interfacial boundary condition, properties of liquids and solids and interactions.\cite{higgins2002timescale, reiter2005residual, damman2007relaxation, peschka2019signatures} Different combinations of short-range (spreading coefficient) and long-range (Hamaker constant) interactions are used to predict the stable or unstable behavior of thin liquid films.\cite{sharma1993relationship,sharma1993equilibrium} It has been also shown that stable thin liquid films can be destabilized using different external stimulus viz. temperature induced marangoni flow, mechanical vibrations, electric or magnetic forces \cite{mitov1998convection, warner2002dewetting, alvarez2008surface, sterman2017rayleigh, kataoka1999patterning, schaeffer2000electrically, surenjav2009manipulation}. However, if the stability condition can be controlled reversibly, one can further investigate the same properties in a reversible manner.\cite{severin2012reversible}

It has been shown that external electric field can be used to very effectively to manipulate the effective interfacial energy.\cite{quilliet2002investigation} Due to the accumulation of the electric charges at a dielectric interface the interfacial free energy changes, which subsequently modified the wetting properties of the interface. Following this idea, many researchers studied the electric field induced dewetting of thin liquid films, theoretically and experimentally.\cite{herminghaus1999dynamical, schaffer2001electrohydrodynamic, verma2005electric, staicu2006electrowetting, priest2006controlled, sahoo2019reversible} The advantage of studying thin film instability using an electric field is to further investigate the behavior of the instability pattern upon removing the electric field. This would help us understand the dynamic change in the interfacial and material properties during the full cycles (voltage ON and voltage OFF). Few recent studies have demonstrated reversible wetting and dewetting in confined geometry (nanopores, microfluidics) using electrowetting.\cite{powell2011electric, li2019ionic} John et al. demonstrated wettability rachets driven fluid transport by switching ON and OFF the applied voltage.\cite{john2007liquid, john2008ratchet}. It is also found that the process of dewetting and spreading do not completely follow the same dynamics. During dewetting, rim of a liquid film recedes at a constant speed with a fixed dynamic contact angle, which quickly relaxes to a spherical shape drop towards the end. On the other hand, during spreading, the contact angle of a drop decreases continuously while the drop maintains its spherical shape.\cite{edwards2016not, edwards2020viscous} 

Thin lubricating films of \textit{Nepenthes'} pitcher plant inspired slippery surfaces are among the most suitable candidates to study the stability of thin liquid films.\cite{lafuma2011slippery, wong2011bioinspired, rao2021highly, epstein2012liquid, wang2016bioinspired, liu2020robust, stamatopoulos2017exceptional} It has been shown that on hydrophilic solid surfaces, the thin lubricating oil films dewet under aqueous drops, which subsequently brings the drops into direct contact with the solid surface and get pinned.\cite{lafuma2011slippery, carlson2013short} Dewetting mechanism and dynamics dominantly depend on the surface and interfacial energies of various components of the system, i.e. solid surface, lubricating fluid, and test liquid.\cite{daniel2017oleoplaning, sharma2019sink, bhatt2022dewetting} In this article, we demonstrate the electric field controlled reversible dewetting of thin lubricating films underneath aqueous drops on stable slippery surfaces over multiple cycles. 

\section{Results and discussion}
\begin{figure*}[t!]
	\centering
		\includegraphics[width=1\textwidth]{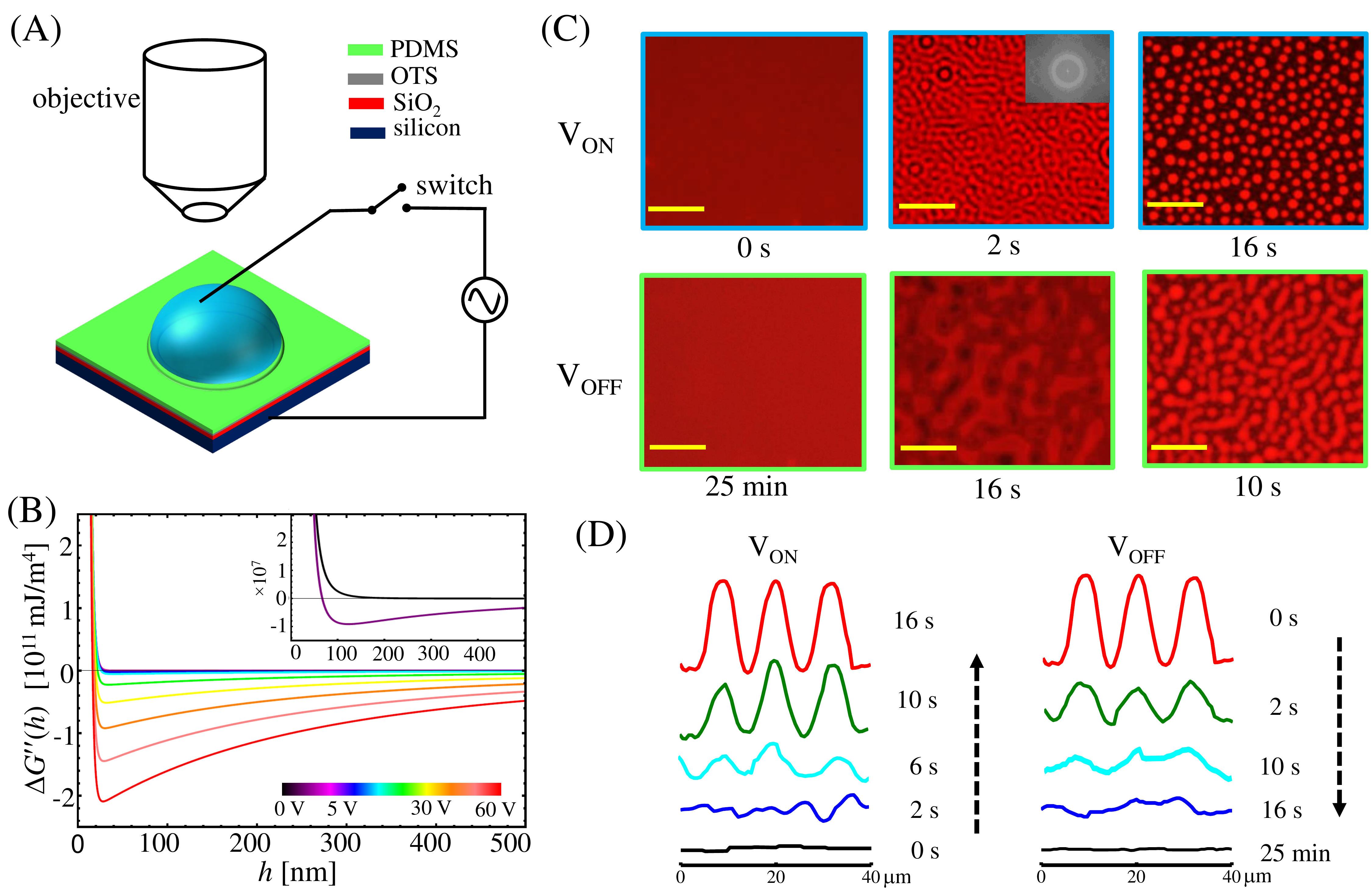}
	\caption{(A) Schematic of the experimental setup. (B) Second derivative of the excess free energy $\Delta G^{\prime \prime}(h)$ showing the unstable behavior of thin PDMS films underneath a drop at different applied voltages. Inset shows the same plot for 0 V and 0.5 V, confirming that thin PDMS films become unstable even at 0.5 V. (C) Fluorescence micrographs of dewetting of thin PDMS films for the forward (after applying 40 V) and reverse (after switching off the voltage) cycles. The scale bar for all the micrographs is 100$\,\upmu$m. (D) The intensity profiles of dewetting dynamics showing the change in the amplitude of the surface waves for the forward (voltage ON) and reverse (voltage OFF) cycles.}
	\label{Fig.1}
\end{figure*}

Figure \ref{Fig.1}(A) shows the schematics of the experimental system where an aqueous drop is deposited on a thin lubricating film (PDMS) coated hydrophobic solid surface, and a potential difference is applied across the drop and the underneath conducting silicon substrate. The experimental system correspond to a stable slippery surface where the lubricating PDMS film provide frictionless slippery interface to the top aqueous drops. The stable lubricating film can be made unstable upon applying an external electric field across the dielectric lubricating film. 
Theoretically, the stability of a thin liquid film can be defined in terms of the total excess free energy of the film, which is the sum of the van der Waals interaction, the acid-base interaction and the Born repulsion.\cite{verma2005electric} In the presence of an external electric field, an additional term due to the electrostatic interaction is added to the total free energy. The electrostatic term is destabilizing in nature and decreases the total free energy of the system. For the present experimental system, Si/SiO$_{2}$/OTS/PDMS/drop, the total excess free energy can be defined as, 
\begin{equation}
\begin{split}
\label{eq:1}
\Delta G(h)\,=-\frac{A_{\mathrm{OTS/PDMS/drop}}}{12 \pi h^{2}}+\frac{A_{\mathrm{OTS/PDMS/drop}}-{A_{\mathrm{SiO_2/PDMS/drop}}}}{12 \pi (\,{h+d_{\mathrm{OTS}}})\,^{2}}\\
+\frac{A_{\mathrm{SiO_2/PDMS/drop}}-A_{\mathrm{Si/PDMS/drop}}}{12 \pi (\,{h+d_{\mathrm{OTS}}+d_{\mathrm{SiO_2}}})\,^{2}}\,+ S_{\mathrm{P}}\,\mathrm{exp}(\,\frac{{d_{\mathrm{min}}}-h}{l})\\
-\frac{1}{2}\frac{\epsilon_\mathrm{0}\epsilon_\mathrm{SiO_2}}{d_\mathrm{SiO_2}}\frac{1}{(1+\frac{d_\mathrm{OTS}\epsilon_\mathrm{SiO_2}}{d_\mathrm{SiO_2}\epsilon_\mathrm{OTS}}+\frac{h\epsilon_\mathrm{SiO_2}}{d_\mathrm{SiO_2}\epsilon_\mathrm{PDMS}})}V^2 + \frac{c_\mathrm{OTS}}{h^8}
\end{split}
\end{equation}
where, $h$, $d_\mathrm{OTS}$, $d_\mathrm{SiO_{2}}$, $d_\mathrm{min}$ and $l$ represent the thickness of PDMS film, OTS monolayer, dielectric SiO$_2$ layer, atomic cut-off distance and polymer correlation length, respectively. $A$ is the effective Hamaker constant for a three-layer system, $S_{\mathrm{P}}$ is the polar component of the spreading coefficient (which determines the acid/base interaction), $c$ is the strength of the short-range interaction and $V$ is the applied $\mathit{rms}$ voltage. Figure \ref{Fig.1}(B) shows the plot of the second derivative of the total excess free energy $(\Delta G^{\prime \prime}(h))$ confirming that thin PDMS films become unstable with applied voltage. With increase in the value of the applied voltage, the magnitude of $\Delta G^{\prime \prime}(h)$ also increases, resulting in the faster dewetting of the thin PDMS films. Inset of Figure \ref{Fig.1}(B) shows the $\Delta G^{\prime \prime}(h)$ curve for 0 V and 0.5 V, which indicates that even such a small voltage can make PDMS films unstable. The total excess free energy equation (Eq. \ref{eq:1}) also suggests that upon removing the applied voltage, thin uniform film will be the stable configuration. 

Figure \ref{Fig.1}(C) top row shows fluorescence images for the dewetting of a 500 nm thin PDMS film underneath an aqueous drop at 40 V (forward cycle). The uniform intensity of the first image (at 0 s) confirms the stable nature of PDMS films at 0 V. After applying the voltage, surface capillary waves appear in milliseconds in random directions, which are similar to the perturbations in spinodal dewetting due to the disjoining pressure \cite{verma2005electric}. The amplitude of these surface waves grows in time, leading to the final dewetting pattern of smaller droplets after about 16 s. The dewetted droplets are found stable as long as the applied voltage is present. Upon reducing the applied voltage to 0 V (reverse cycle), the contact angle of dewetted droplets start decreasing and they start coalescing with neighboring droplets within few seconds. Contact angle of coalesced droplets continue to decrease as they rewet to form a uniform continuous film, similar to the starting one. Here one should note that for 40 V, the total time corresponding to the dewetting in forward cycle is about 16 s, which increases to about 25 mins for complete rewetting in the reverse cycle. Figure \ref{Fig.1}(D) shows the scanline profiles of the surface of a PDMS film for the same times as in the fluorescence images for the forward and reverse voltage cycles. It is clear from the figure that in forward cycle, the perturbations on the smooth surface of a PDMS film grow leading to hole nucleation after about 2 s time, and the final dewetting is complete after about 16 s. On the other hand in the reverse cycle, the final dewetted droplets spread, coalesce and rewet to form a continuous film within 16 s (from the beginning of the reverse cycle) but the film surface contains perturbations. Subsequently, it takes about 25 mins to the film to become completely smooth with no residual surface perturbations. Due to the high viscosity of the PDMS, surface relaxation takes such a long time \cite{edwards2016not, edwards2020viscous}. Also, it is important because we would like to have identical initial condition i.e. completely smooth surface of a PDMS film during different forward cycles. It is also observed that the total time of the reverse cycle (rewetting time) decreases with the increase in the applied potential. For example, at 10 V the dewetting time is 40 mins and the corresponding rewetting time is about 3 h, whereas at 60 V within milliseconds dewetting time, the rewetting time reduces to 6 mins. 

After reducing the voltage to 0 V and formation of uniform PDMS film at the end of the first reserve cycle, another round of dewetting-rewetting cycle was performed to test the repeatability of the system. It was observed that in the second round of dewetting-rewetting cycle, the PDMS film dewets and rewets in a manner identical (with respect to the morphology and time) to the first cycle. The complete reversibility was also observed even in the third dewetting-rewetting cycle. This tells us that thin PDMS films underneath aqueous drops on stable slippery surfaces can reversibly undergo multiple dewetting-rewetting cycles without loosing any of its characteristic feature as the entire process if solely controlled by the applied electric potential. 

\begin{figure*}[t]
	\centering
		\includegraphics[width=1.00\textwidth]{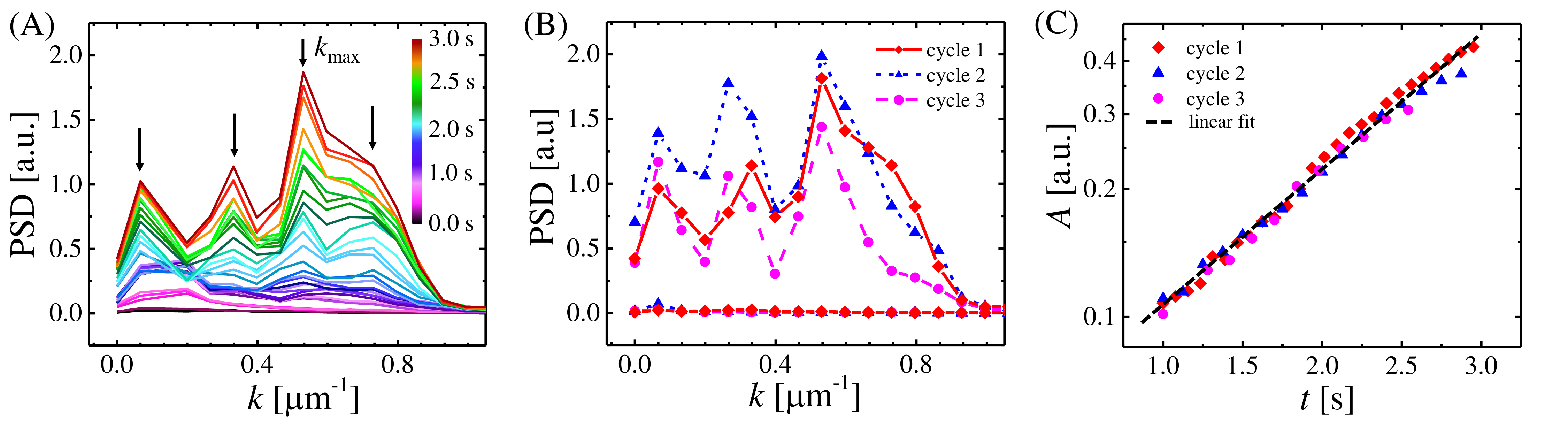}
	\caption{(A) Power spectral density for the forward cycle 1 at 40 V showing temporal evolution and mode selection, resulting in the dominant mode $k_{\mathrm{max}}=0.53\;\upmu\mathrm{m}^{-1}$ along with few other modes at $0.06\;\upmu\mathrm{m}^{-1}, 0.33\;\upmu\mathrm{m}^{-1}\;\&\;0.73\;\upmu\mathrm{m}^{-1}$ $k$ values. (B) Power spectral density at initial and final stages for three consecutive cycles of the dewetting process showing the presence of $k_{\mathrm{max}}$ and other modes. (C) semi-logarithmic plot of the amplitude of the fastest growing mode $k_{\mathrm{max}}$ for the three cycles of the dewetting process.}
	\label{Fig.2}
\end{figure*}

Since the dewetting is being investigated using optical fluorescence microscopy, the temporal evolution of dewetting PDMS films are analyzed in terms of intensity profile, rather than height profile obtained by atomic force microscope of dewetting films \cite{seemann2001gaining, khare2007dewetting}. During dewetting, the capillary waves present on the surface of the a PDMS film under an aqueous drop can be analyzed using the linear stability analysis. Considering the lubricating PDMS fluid as a Newtonian liquid, the Navier-Stokes equation can be simplified using the long-wave approximation. It has been shown that electric field induced instability is a long-wave type where the wavelength of the instability is assumed large compared to the thickness of the film \cite{schaeffer2000electrically}. Since the slip length of PDMS on an OTS grafted surface is about 25 nm, the system can be considered as under weak or moderate slip regime and the linear stability analysis will be similar as for the no-slip regime \cite{scarratt2019slippery, rauscher2008spinodal}. The thin-film equation (spatiotemporal) for the lateral liquid flow under no-slip regime is given by
\begin{equation}
\label{eq:2}
3 \eta \left(\frac{\partial h(x,t)}{\partial t}\right) -\nabla \cdot[h^{3}\nabla P]=0
\end{equation}
where, $\eta$ is the viscosity of PDMS, $h$ is the local film thickness and $P$ is the total pressure across the film due to curvature, disjoining pressure and Maxwell's stress tensor\cite{verma2005electric}. Using the linear stability analysis, Eq. \ref{eq:2} can be linearized with the ansatz $h = h_\mathrm{0} + \epsilon e^{(\omega t-ikx)}$, where $k$ is the wavenumber, $\omega$ is the growth rate, and $\epsilon$ ($<< h_\mathrm{0}$) is the amplitude of surface perturbations. So, the dispersion relation for a capillary wave under long wave approximation can be simplified to
\begin{equation}
\label{eq:3}
\omega = -\frac{h_\mathrm{0}^{3} k^{2}}{3 \eta }\left(\Delta G^{\prime\prime}(h=h_\mathrm{0})+k^{2}\gamma_\mathrm{PD}\right)
\end{equation}
where $\gamma_\mathrm{PD}$ represent the interfacial tension between PDMS and aqueous drop. Since the dispersion relation is in the Fourier space, the Fourier spectrum (Power Spectral Density) of the intensity profile of a dewetting PDMS film at 40 V is shown in Fig. \ref{Fig.2}(A). Since the linear stability analysis is performed in one dimension, smaller area of 100$\times$100 $\upmu \mathrm{m}^2$ was chosen for the analysis to avoid overlapping of multiple capillary waves oriented in different directions. In the beginning at 0 s, the black line in Fig. \ref{Fig.2}(A) confirms that there is no preferred wavelength present in the system. 
After about 800 ms, surface capillary waves with all wavelengths greater than the critical wavelength (corresponding to $\omega$ = 0) appear and start growing, as shown from magenta to red lines in Fig. \ref{Fig.2}(A). Out of these waves, one wave grows with the fastest speed (corresponding to the fastest growing mode) and attempts to suppress the other waves, which correspond to $k_{\mathrm{max}}=0.53;\upmu\mathrm{m}^{-1}$ for the present case. In addition to $k_{\mathrm{max}}$, few other waves at $0.06\;\upmu\mathrm{m}^{-1}, 0.33\;\upmu\mathrm{m}^{-1}\;\&\;0.73\;\upmu\mathrm{m}^{-1}$ $k$ values also grow with relatively slower rate, due to the overlapping of multiple surface waves in different directions (cf. Fig. \ref{Fig.1}(C)). The final red line corresponds to the time when the waves touch the substrate, and holes start nucleating. After this the dewetted holes grow and results in the final dewetted droplets. At the end of the dewetting, the applied voltage was reduced to 0 by switching off the power supply. As a result, the dewetted droplets rewet and coalesce resulting in a homogeneous film at a later stage. Upon applying the same potential of 40 V again, similar temporal evolution of surface waves and mode selection of the dominant wave was observed and shown in Fig. \ref{Fig.2}(B) for three different cycles. Power spectral density in Fig. \ref{Fig.2}(B) shows the red, blue and magenta curves for cycle 1, 2 and 3, respectively at initial and final stages. All characteristic modes, i.e. $k_{\mathrm{max}}$ and other modes, were found to be present in all the three cycles. This further confirms the fully reversible characteristic of the dewetting of thin PDMS films underneath aqueous drops. The mode corresponding to $k_{\mathrm{max}}= 0.53\;\upmu\mathrm{m}^{-1}$ grows with the fastest rate in all three cycles, and its amplitude is plotted in the semi-logarithmic scale in Fig. \ref{Fig.2}(C). So, it is clear that the amplitude of the fastest growing mode $k_{\mathrm{max}}$ grows exponentially in all the three cycles with the time constant $\uptau=3(\pm 0.02)\;\mathrm{s}$. Since the wavenumber of the fastest growing mode remains the same in the three dewetting cycles, it indicates that the mode selection is the inherent property of system and does not depend on the number of cycles.

\begin{figure*}[t!]
	\centering
		\includegraphics[width=0.76\textwidth]{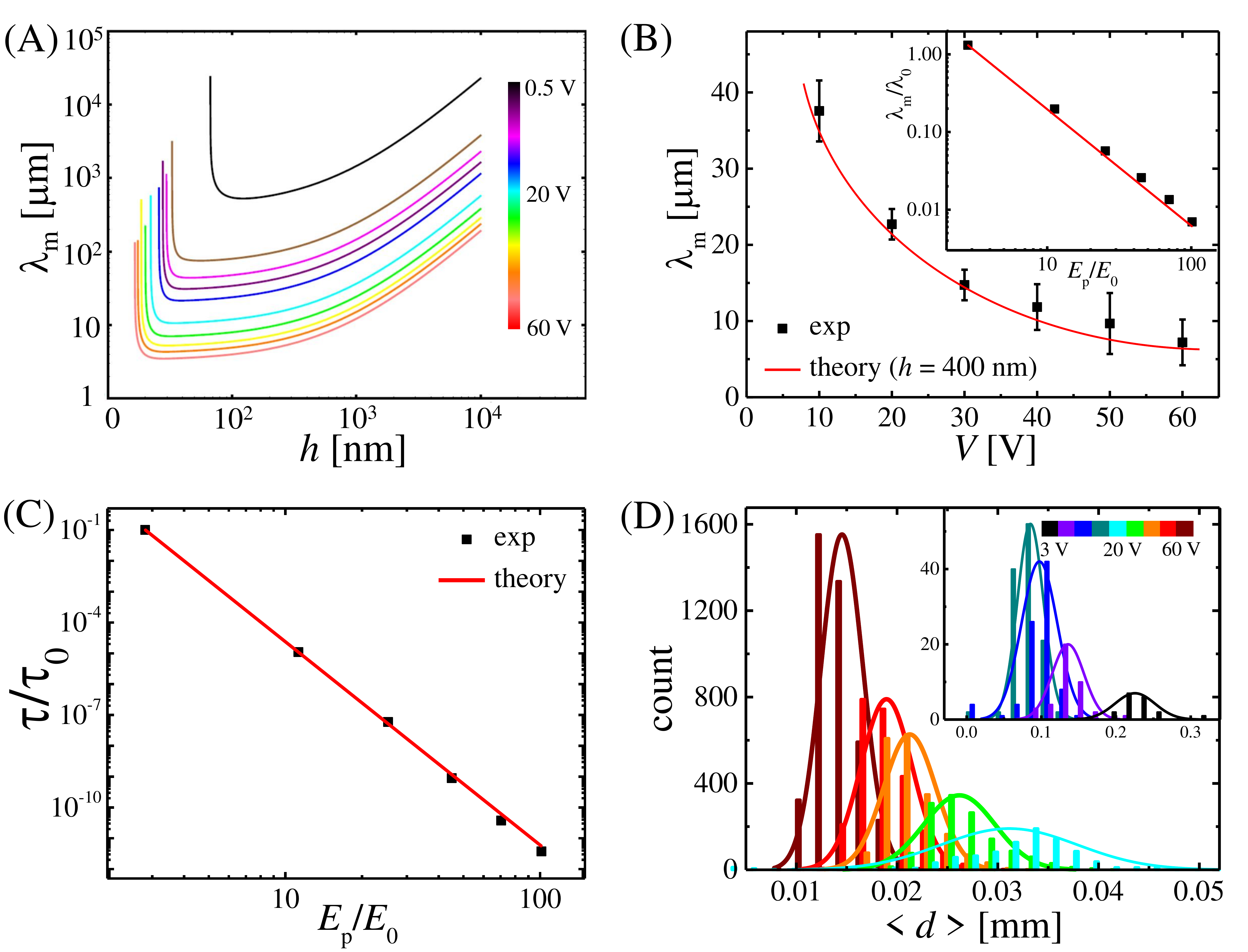}
	\caption{(A) Plot of the wavelength of the fastest growing mode with film thickness using Eq. \ref{eq:4} for voltages from 0.5 V to 60 V. (B) Variation of the wavelength of the fastest growing mode with applied voltage. Black data points represent experimental values and solid red line represents theoretical curve from Eq. \ref{eq:4}. Inset shows rescaled wavelength ($\lambda_\mathrm{m}/\lambda_\mathrm{0}$) with electric field ($E_\mathrm{P}/E_\mathrm{0}$) fitted with the theoretical curve from Eq. \ref{eq:5}. (C) Rescaled plot of the dewetting time constant $\uptau/\uptau_\mathrm{0}$ with the applied electric field ($E_\mathrm{P}/E_\mathrm{0}$) fitted with the theoretical curve from Eq. \ref{eq:7}. (D) The nearest neighbor distance of dewetted PDMS droplets measured from the fluorescence micrographs for different voltages.}
	\label{Fig.3}
\end{figure*}

To verify that the dewetting process follows linear instability analysis, we can also calculate the wavelength of the dominant or fastest growing mode from the dispersion relation (Eq. \ref{eq:3}), using $\mathrm{d} \omega/\mathrm{d} k=0$,
\begin{equation}
\label{eq:4}
\lambda_\mathrm{m} = 2 \pi \left(-\frac{\Delta G^{\prime\prime}(h=h_\mathrm{0})}{2\gamma_\mathrm{PD}}\right)^{-1/2}
\end{equation}
which depends on $\Delta G^{\prime\prime}(h=h_\mathrm{0})$ which subsequently depends on the applied voltage (cf. Eq. \ref{eq:1}). Figure \ref{Fig.3}(A) shows the plot of the wavelength of the fastest growing mode ($\lambda_\mathrm{m}$) with thickness of PDMS films for different applied voltages. The plot tells that for a fixed thickness of PDMS film, $\lambda_\mathrm{m}$ decreases with increasing voltage. Figure \ref{Fig.3}(A) also tells that PDMS films with thickness smaller that $h_{\mathrm{min}}$ are stable, hence will be present between dewetted droplets after the dewetting completes. 
Experimental values of $\lambda_\mathrm{m}$ are obtained from the corresponding Fourier spectrum and plotted with the corresponding theoretical curve (for $h = 400\;\mathrm{nm}$) from Eq. \ref{eq:4} in Fig. \ref{Fig.3}(B). An excellent agreement of the two indicate that the thickness of PDMS films is reduced from 500 nm (as prepared after spin coating) to 400($\pm 100$) nm due to hydrodynamic squeezing after depositing aqueous drops \cite{bhatt2022dewetting, daniel2017oleoplaning}, which finally undergo dewetting. At a voltage smaller than 10 V, the dewetting time becomes very large and the value of $\lambda_\mathrm{m}$ also diverges, so performing the dewetting dynamics experiments becomes challenging. Also, the thickness of stable PDMS film between dewetted droplets increases at smaller voltage, and it adds to the difficulty in analyzing the dewetting dynamics. To generalize the fastest growing wavelength ($\lambda_\mathrm{m}$) with the applied electric field, Eq. \ref{eq:4} can be written in the non-dimensional form as \cite{schaffer2001electrohydrodynamic},
\begin{equation}
\label{eq:5}
\frac{\lambda_\mathrm{m}}{\lambda_\mathrm{0}}=2\pi\left(\frac{E_\mathrm{p}}{E_\mathrm{0}}\right)^{-3/2}
\end{equation}
where, $\lambda_\mathrm{0} = \varepsilon_\mathrm{0}\varepsilon_\mathrm{SiO_2}^3\varepsilon_\mathrm{OTS}^3\varepsilon_\mathrm{PDMS}\mathrm{V}^2/2\gamma_\mathrm{PD}$ is a characteristic wavelength and $E_0 = \mathrm{V}/\lambda_\mathrm{0}$. $E_{p}$ is the net electric field across the entire dielectric layers and can be written as,
\begin{equation}
\label{eq:6}
E_{p}=\frac{\mathrm{V}}{d_\mathrm{PDMS}\varepsilon_\mathrm{OTS}\varepsilon_\mathrm{SiO_2}+d_\mathrm{OTS}\varepsilon_\mathrm{PDMS}\varepsilon_\mathrm{SiO_2}+d_\mathrm{SiO_2}\varepsilon_\mathrm{PDMS}\varepsilon_\mathrm{OTS}}
\end{equation} 
For the present experimental system, $E_{p} \sim 10^6$ V/m. Inset in Fig. \ref{Fig.3}(B) shows the rescaled plot of Eq. \ref{eq:6}, with the solid red line correspond to the theoretical fit with slope -1.5. This again confirms the excellent agreement between the experiments and the theory. 

We also calculated the time constant ($\uptau$) of the instability from the exponential growth of the amplitude of surface waves.\cite{khare2007dewetting} Figure \ref{Fig.3}(C) shows the logarithmic plot of rescaled time constant ($\uptau / \uptau_0$) with the rescaled electric field ($E_P / E_0$). The solid red line represent the theoretical curve obtained from the dispersion relation 
\begin{equation}
\label{eq:7}
\frac{\uptau}{\uptau_\mathrm{0}}=\pi^{4}\left(\frac{E_{p}}{E_\mathrm{0}}\right)^{-6}
\end{equation} 
Therefore the linear stability analysis is able to capture and explain the entire dewetting process of thin PDMS films underneath aqueous drops at different voltages for different cycles. Therefore, as mentioned earlier, the electric field controlled dewetting of 400 nm thick PDMS films is qualitatively similar to the spinodal dewetting of few nanometer thick films controlled only by the van der Waals interaction. 

Plot of the nearest neighbor distance ($<d>$) of PDMS droplets after complete dewetting for different voltages is shown in Fig. \ref{Fig.3}(D). It is clear that the nearest neighbor distance decreases with increasing voltage, as also predicted by Eq. \ref{eq:4}. Qualitatively, $<d>$ follows the same behavior as $\lambda_m$ shown in Fig. \ref{Fig.3}(B) but quantitatively $<d>$ is always slightly larger than $\lambda_m$. This is primarily due to the presence of other waves in addition to the $k_\mathrm{max}$. As a result, after the complete dewetting, $<d>$ corresponding to $k_\mathrm{max}$ and other modes are present in the system, hence it is slightly different compared to the $\lambda_\mathrm{m}$. Also, with increasing voltage, the size of dewetted droplets become smaller hence the number of dewetted droplets increases.

\begin{figure*}[ht!]
	\centering
		\includegraphics[width=0.5\textwidth]{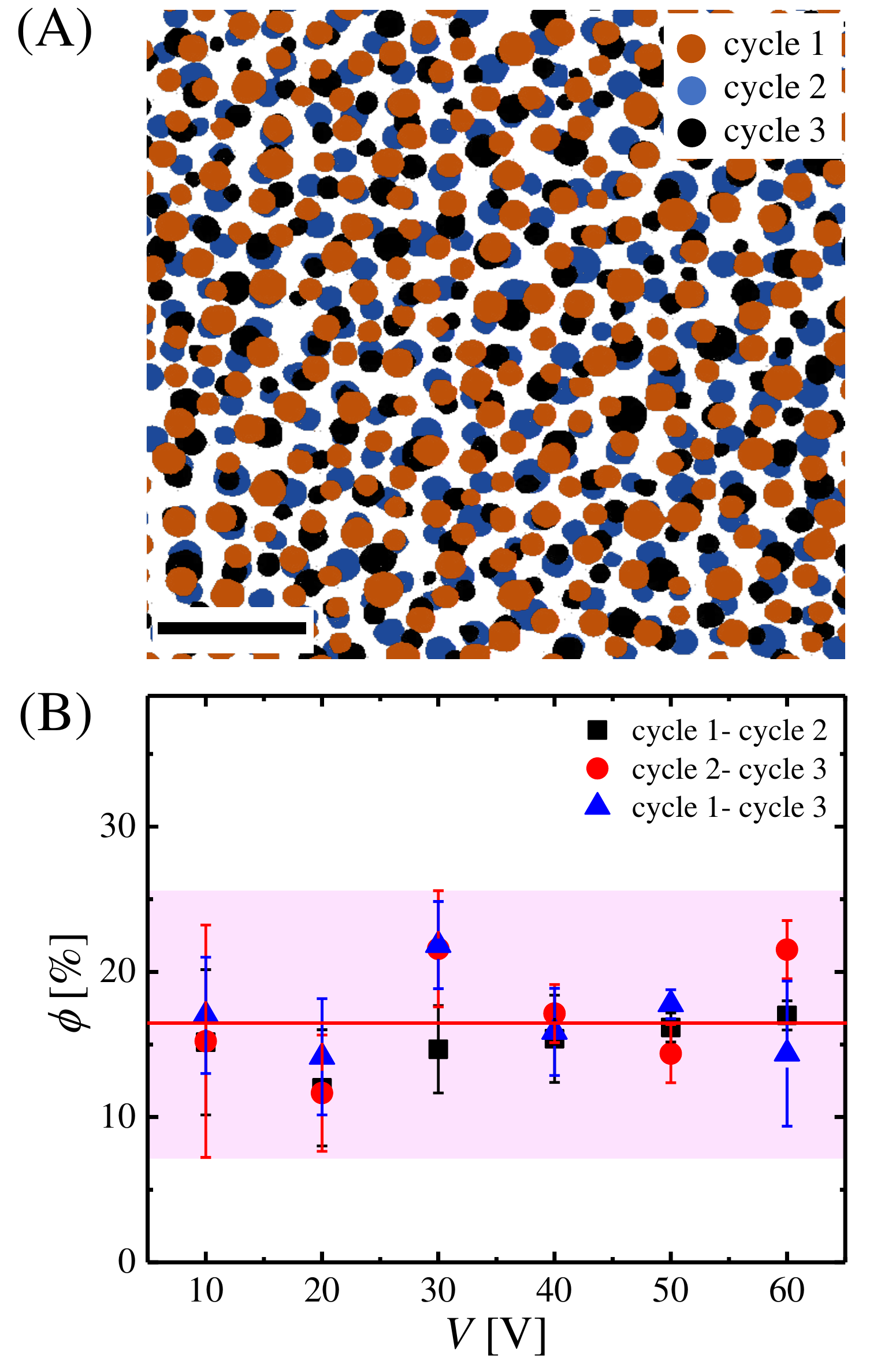}
	\caption{(A) Binary images of the final dewetted droplets for cycle 1, 2 and 3 at 40 V to analyze the correlation of droplet position over different cycles. Different colors is used only to differentiate between different cycles (scale bar 100 $\upmu$m). (B) Overlap area fraction ($\phi$) of the dewetted droplets between cycle 1, cycle 2 and cycle 3 for different voltages. The red line correspond to the mean overlap area fraction at about 16$\%$.}
	\label{Fig.4}
\end{figure*}
Although the dewetting of thin PDMS films underneath aqueous drops with external electric field is fully reversible, it is interesting to analyze the memory effect in the dewetting process over multiple dewetting cycles. Here the memory effect refer to the formation of dewetted PDMS droplets at the same position over different dewetting cycles. As discussed earlier, before the beginning of a new dewetting cycle, it was made sure that the dewetted PDMS film of the previous cycle becomes completely uniform. Memory effect in the dewetting process was analyzed by comparing the positions of final dewetted droplets over multiple dewetting cycles. During image analysis, we overlapped the binary images of dewetted PDMS droplets for different cycles to calculate the area fraction of overlap region of dewetted droplets ($\phi$). Fig. \ref{Fig.4}(A) shows the overlapping of the binary images of dewetted PDMS droplets at 40 V for the cycles 1, 2 \& 3. It is clear from the figure that the dewetted droplets are not located at identical positions over different dewetting cycles, hence the overlap area fraction seems to be quite small. Figure \ref{Fig.4}(B) shows the percentage overlap area fraction at different voltages for the three consecutive cycles. We see that the overlap area fraction between two cycles for different applied voltages is always in the range of 7$\%$ to 26$\%$, with mean around 16$\%$. Since the dewetting process of the thin PDMS films is due to surface capillary waves oriented in random directions, the final dewetted droplets are not expected to have any correlation over different cycles, hence show minimum memory effect. If a dewetting process is dominated by heterogeneous nucleation due to surface defects, dewetted droplets would depict large overlap area fraction showing considerable memory effect in the system.  

\section{Conclusion}
To summarize, the current approach to reversibly control dewetting of thin lubricating films underneath aqueous drops with external electric field can be used to investigate dewetting and rewetting processes in details over multiple cycles. The study shows that stable thin lubricating films underneath aqueous drops on a hydrophobic solid surface dewet upon applying AC voltage across the films. The dewetting process is identical in nature to spinodal dewetting having surface capillary waves of wavelengths greater than the critical wavelength which grow exponentially with time leading to the final pattern with multiple dewetted droplets. The wavelength of the fastest growing mode correspond to the final droplet separation. We observed that even a small voltage of 0.5 V is able to destabilize the lubricating films resulting in smaller number of large sized dewetted droplets. At larger voltages, the minimum of the total free energy increases, hence the films dewet faster with larger number of small sized dewetted droplets. Experimentally obtained droplet separation and dewetting time follow a universal scaling behavior obtained by the linear stability analysis. Upon reducing the applied voltage to 0, the dewetted droplets coalesce and form a uniform film again. Total time of dewetting in a forward cycle is found to be much smaller compared to rewetting in reverse cycle. This asymmetry is due to slow surface relaxation of high viscosity lubricating fluid during rewetting process. The final dewetting pattern, i.e. the distribution of dewetted droplets, do not show any memory effect over multiple dewetting cycles. 
 
\section{Materials and methods}
\textbf{Materials} 
p-type silicon (Si) wafers ($<$100$>$, resistivity 0.001-0.005 ohm-cm) with 1 $\upmu$m thermal oxide layer (University Wafers Inc.) were used as a substrate which serves as the bottom electrode. To prepare stable slippery surfaces, the surface energy of Si wafers were modified by grafting self-assembled monolayer (SAM) of octadecyltrichlorosilane (OTS) (Sigma-Aldrich) molecules. Polydimethylsiloxane (PDMS) (viscosity 5000 cSt, surface tension 21.2 mN/m, Dow Corning) was used as a lubricating fluid for all experiments. Nile red dye, ($\mathrm{C}_{20}\mathrm{H}_{18}\mathrm{N}_{2}\mathrm{O}_{2}$, Sigma-Aldrich) was added in PDMS for fluorescence imaging. Mixture of 80$\%$Glycerol (Fisher Scientific) and 20$\%$ DI water (with 0.1 M NaCl) was used as aqueous liquid (conductivity 0.11 S/m). NaCl is added to enhance the conductivity of the aqueous liquid. The aqueous liquid is hygroscopically stable at our experimental conditions. Copper wire (diameter 70 $\upmu\mathrm{m}$) was connected with the bottom substrate using a silver paste (Fisher Scientific) and a platinum wire (diameter 250 $\upmu\mathrm{m}$) was used as top electrode. 

\textbf{Method} 
Si wafers were cleaned using ethanol, acetone and toluene in ultrasonic bath followed by plasma cleaning (Harrick Plasma) with O$_2$ plasma for 5 min. The cleaned substrates were then immersed in a 0.2 V/V$\%$ of OTS solution in toluene. After taking out, the substrates were rinsed thoroughly with toluene to remove excess non-grafted OTS molecules. Subsequently, the substrated were heated at 90$^\circ$C for 30 mins. To prepare lubricating films, PDMS was diluted with n-heptane (having Nile red dye with 0.0015 W/V$\%$) with 4 W/V$\%$ ratio and was spin-coated for 100 s with 2000 RPM, 10 s acceleration. This resulted in the thickness of lubricating film of $500\;(\;\pm\;10)$ nm. 

\textbf{Experimental setup} 
Thickness of lubricating films was measured using an optical profiler (F-20, KLA USA). Aqueous drop (glycerol-water mixture) of 0.5 ml volume was deposited on the lubricated substrates and left for 15 mins. AC voltage of 1 KHz frequency was taken from a function generator (SG1610C, Aplab India) and amplified using a high voltage amplifier (T-50, Elbatech Italy). A digital oscilloscope (GDS-1062, Gwinstek India) was used to measure the frequency and voltage of the input AC signal. A fluorescence microscope (BX-51, Olympus Japan) equipped with a color CMOS camera (10 fps, $1024\;\mathrm{pixel}\;\times\;798\;\mathrm{pixel}$) was used to observe the dewetting dynamics of lubricating films under aqueous drops. 

\textbf{Image analysis} 
Open-source softwares, Gwyddion and Image J, were used to analyze the fluorescence images of dewetting of lubricating films. Intensity profiles, droplet count, area fraction, power spectral density were calculated using these softwares. 

\bibliography{Manuscript}

\providecommand{\noopsort}[1]{}\providecommand{\singleletter}[1]{#1}%
\providecommand{\latin}[1]{#1}
\makeatletter
\providecommand{\doi}
  {\begingroup\let\do\@makeother\dospecials
  \catcode`\{=1 \catcode`\}=2 \doi@aux}
\providecommand{\doi@aux}[1]{\endgroup\texttt{#1}}
\makeatother
\providecommand*\mcitethebibliography{\thebibliography}
\csname @ifundefined\endcsname{endmcitethebibliography}
  {\let\endmcitethebibliography\endthebibliography}{}
\begin{mcitethebibliography}{44}
\providecommand*\natexlab[1]{#1}
\providecommand*\mciteSetBstSublistMode[1]{}
\providecommand*\mciteSetBstMaxWidthForm[2]{}
\providecommand*\mciteBstWouldAddEndPuncttrue
  {\def\EndOfBibitem{\unskip.}}
\providecommand*\mciteBstWouldAddEndPunctfalse
  {\let\EndOfBibitem\relax}
\providecommand*\mciteSetBstMidEndSepPunct[3]{}
\providecommand*\mciteSetBstSublistLabelBeginEnd[3]{}
\providecommand*\EndOfBibitem{}
\mciteSetBstSublistMode{f}
\mciteSetBstMaxWidthForm{subitem}{(\alph{mcitesubitemcount})}
\mciteSetBstSublistLabelBeginEnd
  {\mcitemaxwidthsubitemform\space}
  {\relax}
  {\relax}

\bibitem[Erdemir(2005)]{erdemir2005review}
Erdemir,~A. Review of engineered tribological interfaces for improved boundary
  lubrication. \emph{Tribology International} \textbf{2005}, \emph{38},
  249--256\relax
\mciteBstWouldAddEndPuncttrue
\mciteSetBstMidEndSepPunct{\mcitedefaultmidpunct}
{\mcitedefaultendpunct}{\mcitedefaultseppunct}\relax
\EndOfBibitem
\bibitem[Higgins \latin{et~al.}(2002)Higgins, Sferrazza, Jones, Jukes, Sharp,
  Dryden, and Webster]{higgins2002timescale}
Higgins,~A.; Sferrazza,~M.; Jones,~R. A.~L.; Jukes,~P.; Sharp,~J.; Dryden,~L.;
  Webster,~J. The timescale of spinodal dewetting at a polymer/polymer
  interface. \emph{The European Physical Journal E} \textbf{2002}, \emph{8},
  137--143\relax
\mciteBstWouldAddEndPuncttrue
\mciteSetBstMidEndSepPunct{\mcitedefaultmidpunct}
{\mcitedefaultendpunct}{\mcitedefaultseppunct}\relax
\EndOfBibitem
\bibitem[Reiter \latin{et~al.}(2005)Reiter, Hamieh, Damman, Sclavons, Gabriele,
  Vilmin, and Rapha{\"e}l]{reiter2005residual}
Reiter,~G.; Hamieh,~M.; Damman,~P.; Sclavons,~S.; Gabriele,~S.; Vilmin,~T.;
  Rapha{\"e}l,~E. Residual stresses in thin polymer films cause rupture and
  dominate early stages of dewetting. \emph{Nature materials} \textbf{2005},
  \emph{4}, 754--758\relax
\mciteBstWouldAddEndPuncttrue
\mciteSetBstMidEndSepPunct{\mcitedefaultmidpunct}
{\mcitedefaultendpunct}{\mcitedefaultseppunct}\relax
\EndOfBibitem
\bibitem[Damman \latin{et~al.}(2007)Damman, Gabriele, Copp{\'e}e, Desprez,
  Villers, Vilmin, Rapha{\"e}l, Hamieh, Al~Akhrass, and
  Reiter]{damman2007relaxation}
Damman,~P.; Gabriele,~S.; Copp{\'e}e,~S.; Desprez,~S.; Villers,~D.; Vilmin,~T.;
  Rapha{\"e}l,~E.; Hamieh,~M.; Al~Akhrass,~S.; Reiter,~G. Relaxation of
  residual stress and reentanglement of polymers in spin-coated films.
  \emph{Physical review letters} \textbf{2007}, \emph{99}, 036101\relax
\mciteBstWouldAddEndPuncttrue
\mciteSetBstMidEndSepPunct{\mcitedefaultmidpunct}
{\mcitedefaultendpunct}{\mcitedefaultseppunct}\relax
\EndOfBibitem
\bibitem[Peschka \latin{et~al.}(2019)Peschka, Haefner, Marquant, Jacobs,
  M{\"u}nch, and Wagner]{peschka2019signatures}
Peschka,~D.; Haefner,~S.; Marquant,~L.; Jacobs,~K.; M{\"u}nch,~A.; Wagner,~B.
  Signatures of slip in dewetting polymer films. \emph{Proceedings of the
  National Academy of Sciences} \textbf{2019}, \emph{116}, 9275--9284\relax
\mciteBstWouldAddEndPuncttrue
\mciteSetBstMidEndSepPunct{\mcitedefaultmidpunct}
{\mcitedefaultendpunct}{\mcitedefaultseppunct}\relax
\EndOfBibitem
\bibitem[Sharma(1993)]{sharma1993relationship}
Sharma,~A. Relationship of thin film stability and morphology to macroscopic
  parameters of wetting in the apolar and polar systems. \emph{Langmuir}
  \textbf{1993}, \emph{9}, 861--869\relax
\mciteBstWouldAddEndPuncttrue
\mciteSetBstMidEndSepPunct{\mcitedefaultmidpunct}
{\mcitedefaultendpunct}{\mcitedefaultseppunct}\relax
\EndOfBibitem
\bibitem[Sharma(1993)]{sharma1993equilibrium}
Sharma,~A. Equilibrium contact angles and film thicknesses in the apolar and
  polar systems: Role of intermolecular interactions in coexistence of drops
  with thin films. \emph{Langmuir} \textbf{1993}, \emph{9}, 3580--3586\relax
\mciteBstWouldAddEndPuncttrue
\mciteSetBstMidEndSepPunct{\mcitedefaultmidpunct}
{\mcitedefaultendpunct}{\mcitedefaultseppunct}\relax
\EndOfBibitem
\bibitem[Mitov and Kumacheva(1998)Mitov, and Kumacheva]{mitov1998convection}
Mitov,~Z.; Kumacheva,~E. Convection-induced patterns in phase-separating
  polymeric fluids. \emph{Physical Review Letters} \textbf{1998}, \emph{81},
  3427\relax
\mciteBstWouldAddEndPuncttrue
\mciteSetBstMidEndSepPunct{\mcitedefaultmidpunct}
{\mcitedefaultendpunct}{\mcitedefaultseppunct}\relax
\EndOfBibitem
\bibitem[Warner \latin{et~al.}(2002)Warner, Craster, and
  Matar]{warner2002dewetting}
Warner,~M.; Craster,~R.; Matar,~O. Dewetting of ultrathin surfactant-covered
  films. \emph{Physics of Fluids} \textbf{2002}, \emph{14}, 4040--4054\relax
\mciteBstWouldAddEndPuncttrue
\mciteSetBstMidEndSepPunct{\mcitedefaultmidpunct}
{\mcitedefaultendpunct}{\mcitedefaultseppunct}\relax
\EndOfBibitem
\bibitem[Alvarez \latin{et~al.}(2008)Alvarez, Friend, and
  Yeo]{alvarez2008surface}
Alvarez,~M.; Friend,~J.~R.; Yeo,~L.~Y. Surface vibration induced spatial
  ordering of periodic polymer patterns on a substrate. \emph{Langmuir}
  \textbf{2008}, \emph{24}, 10629--10632\relax
\mciteBstWouldAddEndPuncttrue
\mciteSetBstMidEndSepPunct{\mcitedefaultmidpunct}
{\mcitedefaultendpunct}{\mcitedefaultseppunct}\relax
\EndOfBibitem
\bibitem[Sterman-Cohen \latin{et~al.}(2017)Sterman-Cohen, Bestehorn, and
  Oron]{sterman2017rayleigh}
Sterman-Cohen,~E.; Bestehorn,~M.; Oron,~A. Rayleigh-Taylor instability in thin
  liquid films subjected to harmonic vibration. \emph{Physics of Fluids}
  \textbf{2017}, \emph{29}, 052105\relax
\mciteBstWouldAddEndPuncttrue
\mciteSetBstMidEndSepPunct{\mcitedefaultmidpunct}
{\mcitedefaultendpunct}{\mcitedefaultseppunct}\relax
\EndOfBibitem
\bibitem[Kataoka and Troian(1999)Kataoka, and Troian]{kataoka1999patterning}
Kataoka,~D.~E.; Troian,~S.~M. Patterning liquid flow on the microscopic scale.
  \emph{Nature} \textbf{1999}, \emph{402}, 794--797\relax
\mciteBstWouldAddEndPuncttrue
\mciteSetBstMidEndSepPunct{\mcitedefaultmidpunct}
{\mcitedefaultendpunct}{\mcitedefaultseppunct}\relax
\EndOfBibitem
\bibitem[Schaeffer \latin{et~al.}(2000)Schaeffer, Thurn-Albrecht, Russell, and
  Steiner]{schaeffer2000electrically}
Schaeffer,~E.; Thurn-Albrecht,~T.; Russell,~T.~P.; Steiner,~U. Electrically
  induced structure formation and pattern transfer. \emph{Nature}
  \textbf{2000}, \emph{403}, 874--877\relax
\mciteBstWouldAddEndPuncttrue
\mciteSetBstMidEndSepPunct{\mcitedefaultmidpunct}
{\mcitedefaultendpunct}{\mcitedefaultseppunct}\relax
\EndOfBibitem
\bibitem[Surenjav \latin{et~al.}(2009)Surenjav, Priest, Herminghaus, and
  Seemann]{surenjav2009manipulation}
Surenjav,~E.; Priest,~C.; Herminghaus,~S.; Seemann,~R. Manipulation of gel
  emulsions by variable microchannel geometry. \emph{Lab Chip} \textbf{2009},
  \emph{9}, 325--330\relax
\mciteBstWouldAddEndPuncttrue
\mciteSetBstMidEndSepPunct{\mcitedefaultmidpunct}
{\mcitedefaultendpunct}{\mcitedefaultseppunct}\relax
\EndOfBibitem
\bibitem[Severin \latin{et~al.}(2012)Severin, Lange, Sokolov, and
  Rabe]{severin2012reversible}
Severin,~N.; Lange,~P.; Sokolov,~I.~M.; Rabe,~J.~P. Reversible dewetting of a
  molecularly thin fluid water film in a soft graphene--mica slit pore.
  \emph{Nano letters} \textbf{2012}, \emph{12}, 774--779\relax
\mciteBstWouldAddEndPuncttrue
\mciteSetBstMidEndSepPunct{\mcitedefaultmidpunct}
{\mcitedefaultendpunct}{\mcitedefaultseppunct}\relax
\EndOfBibitem
\bibitem[Quilliet and Berge(2002)Quilliet, and
  Berge]{quilliet2002investigation}
Quilliet,~C.; Berge,~B. Investigation of effective interface potentials by
  electrowetting. \emph{EPL (Europhysics Letters)} \textbf{2002}, \emph{60},
  99\relax
\mciteBstWouldAddEndPuncttrue
\mciteSetBstMidEndSepPunct{\mcitedefaultmidpunct}
{\mcitedefaultendpunct}{\mcitedefaultseppunct}\relax
\EndOfBibitem
\bibitem[Herminghaus(1999)]{herminghaus1999dynamical}
Herminghaus,~S. Dynamical instability of thin liquid films between conducting
  media. \emph{Physical review letters} \textbf{1999}, \emph{83}, 2359\relax
\mciteBstWouldAddEndPuncttrue
\mciteSetBstMidEndSepPunct{\mcitedefaultmidpunct}
{\mcitedefaultendpunct}{\mcitedefaultseppunct}\relax
\EndOfBibitem
\bibitem[Sch{\"a}ffer \latin{et~al.}(2001)Sch{\"a}ffer, Thurn-Albrecht,
  Russell, and Steiner]{schaffer2001electrohydrodynamic}
Sch{\"a}ffer,~E.; Thurn-Albrecht,~T.; Russell,~T.~P.; Steiner,~U.
  Electrohydrodynamic instabilities in polymer films. \emph{EPL (Europhysics
  Letters)} \textbf{2001}, \emph{53}, 518\relax
\mciteBstWouldAddEndPuncttrue
\mciteSetBstMidEndSepPunct{\mcitedefaultmidpunct}
{\mcitedefaultendpunct}{\mcitedefaultseppunct}\relax
\EndOfBibitem
\bibitem[Verma \latin{et~al.}(2005)Verma, Sharma, Kargupta, and
  Bhaumik]{verma2005electric}
Verma,~R.; Sharma,~A.; Kargupta,~K.; Bhaumik,~J. Electric field induced
  instability and pattern formation in thin liquid films. \emph{Langmuir}
  \textbf{2005}, \emph{21}, 3710--3721\relax
\mciteBstWouldAddEndPuncttrue
\mciteSetBstMidEndSepPunct{\mcitedefaultmidpunct}
{\mcitedefaultendpunct}{\mcitedefaultseppunct}\relax
\EndOfBibitem
\bibitem[Staicu and Mugele(2006)Staicu, and Mugele]{staicu2006electrowetting}
Staicu,~A.; Mugele,~F. Electrowetting-induced oil film entrapment and
  instability. \emph{Physical review letters} \textbf{2006}, \emph{97},
  167801\relax
\mciteBstWouldAddEndPuncttrue
\mciteSetBstMidEndSepPunct{\mcitedefaultmidpunct}
{\mcitedefaultendpunct}{\mcitedefaultseppunct}\relax
\EndOfBibitem
\bibitem[Priest \latin{et~al.}(2006)Priest, Herminghaus, and
  Seemann]{priest2006controlled}
Priest,~C.; Herminghaus,~S.; Seemann,~R. Controlled electrocoalescence in
  microfluidics: Targeting a single lamella. \emph{Applied Physics Letters}
  \textbf{2006}, \emph{89}, 134101\relax
\mciteBstWouldAddEndPuncttrue
\mciteSetBstMidEndSepPunct{\mcitedefaultmidpunct}
{\mcitedefaultendpunct}{\mcitedefaultseppunct}\relax
\EndOfBibitem
\bibitem[Sahoo \latin{et~al.}(2019)Sahoo, Bhandaru, and
  Mukherjee]{sahoo2019reversible}
Sahoo,~S.; Bhandaru,~N.; Mukherjee,~R. Reversible morphological switching and
  deformation hysteresis in electric field mediated instability of thin elastic
  films. \emph{Soft Matter} \textbf{2019}, \emph{15}, 3828--3834\relax
\mciteBstWouldAddEndPuncttrue
\mciteSetBstMidEndSepPunct{\mcitedefaultmidpunct}
{\mcitedefaultendpunct}{\mcitedefaultseppunct}\relax
\EndOfBibitem
\bibitem[Powell \latin{et~al.}(2011)Powell, Cleary, Davenport, Shea, and
  Siwy]{powell2011electric}
Powell,~M.~R.; Cleary,~L.; Davenport,~M.; Shea,~K.~J.; Siwy,~Z.~S.
  Electric-field-induced wetting and dewetting in single hydrophobic nanopores.
  \emph{Nature nanotechnology} \textbf{2011}, \emph{6}, 798--802\relax
\mciteBstWouldAddEndPuncttrue
\mciteSetBstMidEndSepPunct{\mcitedefaultmidpunct}
{\mcitedefaultendpunct}{\mcitedefaultseppunct}\relax
\EndOfBibitem
\bibitem[Li \latin{et~al.}(2019)Li, Ha, Liu, van Dam, and
  CJ’Kim]{li2019ionic}
Li,~J.; Ha,~N.~S.; Liu,~T.; van Dam,~R.~M.; CJ’Kim,~C.-J.
  Ionic-surfactant-mediated electro-dewetting for digital microfluidics.
  \emph{Nature} \textbf{2019}, \emph{572}, 507--510\relax
\mciteBstWouldAddEndPuncttrue
\mciteSetBstMidEndSepPunct{\mcitedefaultmidpunct}
{\mcitedefaultendpunct}{\mcitedefaultseppunct}\relax
\EndOfBibitem
\bibitem[John and Thiele(2007)John, and Thiele]{john2007liquid}
John,~K.; Thiele,~U. Liquid transport generated by a flashing field-induced
  wettability ratchet. \emph{Applied physics letters} \textbf{2007}, \emph{90},
  264102\relax
\mciteBstWouldAddEndPuncttrue
\mciteSetBstMidEndSepPunct{\mcitedefaultmidpunct}
{\mcitedefaultendpunct}{\mcitedefaultseppunct}\relax
\EndOfBibitem
\bibitem[John \latin{et~al.}(2008)John, H{\"a}nggi, and
  Thiele]{john2008ratchet}
John,~K.; H{\"a}nggi,~P.; Thiele,~U. Ratchet-driven fluid transport in bounded
  two-layer films of immiscible liquids. \emph{Soft Matter} \textbf{2008},
  \emph{4}, 1183--1195\relax
\mciteBstWouldAddEndPuncttrue
\mciteSetBstMidEndSepPunct{\mcitedefaultmidpunct}
{\mcitedefaultendpunct}{\mcitedefaultseppunct}\relax
\EndOfBibitem
\bibitem[Edwards \latin{et~al.}(2016)Edwards, Ledesma-Aguilar, Newton, Brown,
  and McHale]{edwards2016not}
Edwards,~A.~M.; Ledesma-Aguilar,~R.; Newton,~M.~I.; Brown,~C.~V.; McHale,~G.
  Not spreading in reverse: The dewetting of a liquid film into a single drop.
  \emph{Science advances} \textbf{2016}, \emph{2}, e1600183\relax
\mciteBstWouldAddEndPuncttrue
\mciteSetBstMidEndSepPunct{\mcitedefaultmidpunct}
{\mcitedefaultendpunct}{\mcitedefaultseppunct}\relax
\EndOfBibitem
\bibitem[Edwards \latin{et~al.}(2020)Edwards, Ledesma-Aguilar, Newton, Brown,
  and McHale]{edwards2020viscous}
Edwards,~A.; Ledesma-Aguilar,~R.; Newton,~M.~I.; Brown,~C.; McHale,~G. A
  viscous switch for liquid-liquid dewetting. \emph{Communications Physics}
  \textbf{2020}, \emph{3}, 1--6\relax
\mciteBstWouldAddEndPuncttrue
\mciteSetBstMidEndSepPunct{\mcitedefaultmidpunct}
{\mcitedefaultendpunct}{\mcitedefaultseppunct}\relax
\EndOfBibitem
\bibitem[Lafuma and Qu{\'e}r{\'e}(2011)Lafuma, and
  Qu{\'e}r{\'e}]{lafuma2011slippery}
Lafuma,~A.; Qu{\'e}r{\'e},~D. Slippery pre-suffused surfaces. \emph{EPL
  (Europhysics Letters)} \textbf{2011}, \emph{96}, 56001\relax
\mciteBstWouldAddEndPuncttrue
\mciteSetBstMidEndSepPunct{\mcitedefaultmidpunct}
{\mcitedefaultendpunct}{\mcitedefaultseppunct}\relax
\EndOfBibitem
\bibitem[Wong \latin{et~al.}(2011)Wong, Kang, Tang, Smythe, Hatton, Grinthal,
  and Aizenberg]{wong2011bioinspired}
Wong,~T.-S.; Kang,~S.~H.; Tang,~S.~K.; Smythe,~E.~J.; Hatton,~B.~D.;
  Grinthal,~A.; Aizenberg,~J. Bioinspired self-repairing slippery surfaces with
  pressure-stable omniphobicity. \emph{Nature} \textbf{2011}, \emph{477},
  443--447\relax
\mciteBstWouldAddEndPuncttrue
\mciteSetBstMidEndSepPunct{\mcitedefaultmidpunct}
{\mcitedefaultendpunct}{\mcitedefaultseppunct}\relax
\EndOfBibitem
\bibitem[Rao \latin{et~al.}(2021)Rao, Lu, Song, Hou, Zhan, and
  Zhang]{rao2021highly}
Rao,~Q.; Lu,~Y.; Song,~L.; Hou,~Y.; Zhan,~X.; Zhang,~Q. Highly Efficient
  Self-Repairing Slippery Liquid-Infused Surface with Promising Anti-Icing and
  Anti-Fouling Performance. \emph{ACS Applied Materials \& Interfaces}
  \textbf{2021}, \emph{13}, 40032--40041\relax
\mciteBstWouldAddEndPuncttrue
\mciteSetBstMidEndSepPunct{\mcitedefaultmidpunct}
{\mcitedefaultendpunct}{\mcitedefaultseppunct}\relax
\EndOfBibitem
\bibitem[Epstein \latin{et~al.}(2012)Epstein, Wong, Belisle, Boggs, and
  Aizenberg]{epstein2012liquid}
Epstein,~A.~K.; Wong,~T.-S.; Belisle,~R.~A.; Boggs,~E.~M.; Aizenberg,~J.
  Liquid-infused structured surfaces with exceptional anti-biofouling
  performance. \emph{Proceedings of the National Academy of Sciences}
  \textbf{2012}, \emph{109}, 13182--13187\relax
\mciteBstWouldAddEndPuncttrue
\mciteSetBstMidEndSepPunct{\mcitedefaultmidpunct}
{\mcitedefaultendpunct}{\mcitedefaultseppunct}\relax
\EndOfBibitem
\bibitem[Wang \latin{et~al.}(2016)Wang, Kato, Blois, and
  Wong]{wang2016bioinspired}
Wang,~J.; Kato,~K.; Blois,~A.~P.; Wong,~T.-S. Bioinspired omniphobic coatings
  with a thermal self-repair function on industrial materials. \emph{ACS
  applied materials \& interfaces} \textbf{2016}, \emph{8}, 8265--8271\relax
\mciteBstWouldAddEndPuncttrue
\mciteSetBstMidEndSepPunct{\mcitedefaultmidpunct}
{\mcitedefaultendpunct}{\mcitedefaultseppunct}\relax
\EndOfBibitem
\bibitem[Liu \latin{et~al.}(2020)Liu, Li, Lu, Liu, Feng, and
  Liu]{liu2020robust}
Liu,~C.; Li,~Y.; Lu,~C.; Liu,~Y.; Feng,~S.; Liu,~Y. Robust slippery
  liquid-infused porous network surfaces for enhanced anti-icing/deicing
  performance. \emph{ACS applied materials \& interfaces} \textbf{2020},
  \emph{12}, 25471--25477\relax
\mciteBstWouldAddEndPuncttrue
\mciteSetBstMidEndSepPunct{\mcitedefaultmidpunct}
{\mcitedefaultendpunct}{\mcitedefaultseppunct}\relax
\EndOfBibitem
\bibitem[Stamatopoulos \latin{et~al.}(2017)Stamatopoulos, Hemrle, Wang, and
  Poulikakos]{stamatopoulos2017exceptional}
Stamatopoulos,~C.; Hemrle,~J.; Wang,~D.; Poulikakos,~D. Exceptional anti-icing
  performance of self-impregnating slippery surfaces. \emph{ACS applied
  materials \& interfaces} \textbf{2017}, \emph{9}, 10233--10242\relax
\mciteBstWouldAddEndPuncttrue
\mciteSetBstMidEndSepPunct{\mcitedefaultmidpunct}
{\mcitedefaultendpunct}{\mcitedefaultseppunct}\relax
\EndOfBibitem
\bibitem[Carlson \latin{et~al.}(2013)Carlson, Kim, Amberg, and
  Stone]{carlson2013short}
Carlson,~A.; Kim,~P.; Amberg,~G.; Stone,~H.~A. Short and long time drop
  dynamics on lubricated substrates. \emph{EPL} \textbf{2013}, \emph{104},
  34008\relax
\mciteBstWouldAddEndPuncttrue
\mciteSetBstMidEndSepPunct{\mcitedefaultmidpunct}
{\mcitedefaultendpunct}{\mcitedefaultseppunct}\relax
\EndOfBibitem
\bibitem[Daniel \latin{et~al.}(2017)Daniel, Timonen, Li, Velling, and
  Aizenberg]{daniel2017oleoplaning}
Daniel,~D.; Timonen,~J.~V.; Li,~R.; Velling,~S.~J.; Aizenberg,~J. Oleoplaning
  droplets on lubricated surfaces. \emph{Nat. Phys.} \textbf{2017}, \emph{13},
  1020--1025\relax
\mciteBstWouldAddEndPuncttrue
\mciteSetBstMidEndSepPunct{\mcitedefaultmidpunct}
{\mcitedefaultendpunct}{\mcitedefaultseppunct}\relax
\EndOfBibitem
\bibitem[Sharma \latin{et~al.}(2019)Sharma, Roy, Pant, and
  Khare]{sharma2019sink}
Sharma,~M.; Roy,~P.~K.; Pant,~R.; Khare,~K. Sink dynamics of aqueous drops on
  lubricating fluid coated hydrophilic surfaces. \emph{Colloids and Surfaces A:
  Physicochemical and Engineering Aspects} \textbf{2019}, \emph{562},
  377--382\relax
\mciteBstWouldAddEndPuncttrue
\mciteSetBstMidEndSepPunct{\mcitedefaultmidpunct}
{\mcitedefaultendpunct}{\mcitedefaultseppunct}\relax
\EndOfBibitem
\bibitem[Bhatt \latin{et~al.}(2022)Bhatt, Gupta, Sharma, and
  Khare]{bhatt2022dewetting}
Bhatt,~B.; Gupta,~S.; Sharma,~M.; Khare,~K. Dewetting of non-polar thin
  lubricating films underneath polar liquid drops on slippery surfaces.
  \emph{Journal of Colloid and Interface Science} \textbf{2022}, \emph{607},
  530--537\relax
\mciteBstWouldAddEndPuncttrue
\mciteSetBstMidEndSepPunct{\mcitedefaultmidpunct}
{\mcitedefaultendpunct}{\mcitedefaultseppunct}\relax
\EndOfBibitem
\bibitem[Seemann \latin{et~al.}(2001)Seemann, Herminghaus, and
  Jacobs]{seemann2001gaining}
Seemann,~R.; Herminghaus,~S.; Jacobs,~K. Gaining control of pattern formation
  of dewetting liquid films. \emph{Journal of Physics: Condensed Matter}
  \textbf{2001}, \emph{13}, 4925\relax
\mciteBstWouldAddEndPuncttrue
\mciteSetBstMidEndSepPunct{\mcitedefaultmidpunct}
{\mcitedefaultendpunct}{\mcitedefaultseppunct}\relax
\EndOfBibitem
\bibitem[Khare \latin{et~al.}(2007)Khare, Brinkmann, Law, Gurevich,
  Herminghaus, and Seemann]{khare2007dewetting}
Khare,~K.; Brinkmann,~M.; Law,~B.~M.; Gurevich,~E.~L.; Herminghaus,~S.;
  Seemann,~R. Dewetting of liquid filaments in wedge-shaped grooves.
  \emph{Langmuir} \textbf{2007}, \emph{23}, 12138--12141\relax
\mciteBstWouldAddEndPuncttrue
\mciteSetBstMidEndSepPunct{\mcitedefaultmidpunct}
{\mcitedefaultendpunct}{\mcitedefaultseppunct}\relax
\EndOfBibitem
\bibitem[Scarratt \latin{et~al.}(2019)Scarratt, Zhu, and
  Neto]{scarratt2019slippery}
Scarratt,~L.~R.; Zhu,~L.; Neto,~C. How Slippery are SLIPS? Measuring Effective
  Slip on Lubricated Surfaces with Colloidal Probe Atmoc Force Microscopy.
  \emph{Langmuir} \textbf{2019}, \emph{35}, 2976--2982\relax
\mciteBstWouldAddEndPuncttrue
\mciteSetBstMidEndSepPunct{\mcitedefaultmidpunct}
{\mcitedefaultendpunct}{\mcitedefaultseppunct}\relax
\EndOfBibitem
\bibitem[Rauscher \latin{et~al.}(2008)Rauscher, Blossey, Munch, and
  Wagner]{rauscher2008spinodal}
Rauscher,~M.; Blossey,~R.; Munch,~A.; Wagner,~B. Spinodal dewetting of thin
  films with large interfacial slip: implications from the dispersion relation.
  \emph{Langmuir} \textbf{2008}, \emph{24}, 12290--12294\relax
\mciteBstWouldAddEndPuncttrue
\mciteSetBstMidEndSepPunct{\mcitedefaultmidpunct}
{\mcitedefaultendpunct}{\mcitedefaultseppunct}\relax
\EndOfBibitem
\end{mcitethebibliography}

\end{document}